\documentclass[aps,twocolumn]{revtex4}
\usepackage{mathrsfs}
\usepackage{amsmath,amssymb,amsfonts}

\newcommand{\be}{\begin{equation}}
\newcommand{\ee}{\end{equation}}

\begin{document}

\title{Challenges facing holographic models of QCD}

\author{Thomas D. Cohen}
\email{cohen@physics.umd.edu}

\affiliation{Department of Physics, University of Maryland,
College Park, MD 20742-4111}

\begin{abstract}
This paper, written in memory of Manoj K. Banerjee, takes a
critical look at holographic models of QCD focusing on
``practical'' models in which the five-dimensional theory is
treated classically.  A number of theoretical and phenomenological
challenges to the approach are discussed.
\end{abstract}

\maketitle

Manoj K. Banerjee served as a mentor to me during the early stages
of my career; he had a profound influence on my scientific
development.    Without exaggeration, I can say that I owe my
scientific career to him.  Thus, I am deeply honored to have an
opportunity to write a paper in this issue of the Indian Journal
of Physics dedicated to his memory.

I first got to know Manoj in 1985.  I had arrived at the
University of Maryland as a fresh Ph.D. having done a thesis in
nuclear structure physics related to the then trendy Interacting
Boson Model.   I was looking to expand my horizons.  The nuclear
theory group at Maryland has an admirable tradition: postdocs are
not formally assigned to a given faculty member but can work with
whomever they please.  I naturally gravitated towards Manoj.  In
the first place he was doing very interesting science: soliton
models of the nucleon\cite{Manoj_sol}.  Perhaps equally attractive
was the sheer intellectual excitement of his research effort and
the obvious and infectious joy and enthusiasm with which he did
science.

Manoj's enthusiasm could lead to some {\it very} intense
scientific discussions. I recall an incident which characterizes
the intensity of those days. A scientist was visiting our group to
give a seminar on a competing model of the nucleon.  Before his
talk our visitor, along with me, and Wojeich Broniowski (who at
that time was Manoj's Ph.D. student) met in Manoj's office to
discuss this model.  The discussion became rather heated.  Indeed,
it became so heated that an undergraduate attending office hours
in Wally van Orden's office (a couple of offices away) asked, ``Is
everything OK in there? It sounds like they are having a fight.''
It is worth noting though that while Manoj was often heated in his
discussion of science, he was always free of malice.  Rather, the
intensity was driven by a deep need to get to the core of things
along with an incisive mind and critical scientific judgment. This
critical scientific judgment was omnipresent in Manoj.

One aspect of this which I always appreciated was his extreme
reluctance to accept arguments based on appeals to authority. He
was also very suspicious of ideas which seemed to be trendy. He
also had a rather healthy skepticism toward approaches which where
driven by elegant mathematics rather than rooted in the physics.
In this context, it is interesting to contemplate what his
reaction would have been to the recent emergence of so-called
``holographic models of QCD'', an approach which sometimes goes
under the name of
AdS/QCD\cite{EKSS,DRP1,DRP2,Ghoroku,KatzTensorMesons,HirnRuizSanz,ErlichKribsLow,KKSS,BrodskydeTeramond,ShockWu,EvansTedderWaterson,CsakiReece,ShockBackreacted,Schafer,Forkel,KweeLebed,ColangeloEtAl}.
This approach is based on the hope that some truly spectacular
results in mathematical physics arising from issues in string
theory might be somewhat modified and then applied directly to
QCD.  Certainly the approach \emph{is} trendy and \emph{does} seem
driven by the elegant mathematics.  Moreover, one might sense at
least an implied appeal to authority as the underlying ideas are
taken from Maldecena's remarkable insight relating conformal field
theories in four dimensions to anti-deSitter space in five
dimensions---the so-called AdS/CFT
duality\cite{Maldacena,GubserEtAl,Witten}.  My suspicion is that
Manoj would have a number of deep questions about the entire
enterprise of holographic models of QCD.

It seems to me that it is appropriate in this volume dedicated to
the memory of Manoj Banerjee to write a paper which celebrates his
spirit of critically probing the models in the field.  This paper
will take a critical look at a few aspects of what has been called
the bottom-up approach to holographic QCD.  It is not intended as
a review of the literature---which in any event is growing
rapidly.  Rather it is intended to raise a number of questions
about the foundations of the approach which will need to be
addressed before it can be considered as a viable approach to
hadronic physics.  It should be noted that none of the issues
discussed in this paper are particularly subtle and all should be
known at some level to workers in the field.  The purpose of this
paper is to emphasize at least some of the challenges facing these
models and to stress the need for the community to deal with them.

Before proceeding with this discussion, a few words of caution are
probably in order.  I should noted that Manoj was much more
sensible than me.  While in person he was always \emph{very}
outspoken about his intellectual concerns about the foundations of
various models on the market, he had the good sense not to write
articles cataloging them.  This was sensible for a number of
reasons.  One key reason is simply that on occasion concerns about
the foundations of a model are not so much wrong as simply off
base.  This can happen if the aims of a given model are rather
modest.  Then, it may happen that while the objections may be
trenchant, they are only so if the model is taken to apply more
widely than is intended.  For example, Manoj always had a great
skepticism about the Skyrme model\cite{Skyrme}.  Given what I
wrote before, this is hardly surprising: it was trendy, driven as
much by elegant mathematics as by physics and often supported by
appeals to Witten's authority.  Of course the Skyrme model {\it
was} often abused: it was used where it ought not to have
been---and Manoj was always quick to spot this.  However, I always
had the sense that he never fully appreciated that while the model
is quite limited---it could only be sensibly used for a narrow
class of problems---within this limited domain the model makes a
good bit of sense, at least as an illustrative model.  It is quite
likely that a similar situation will arise for the models here.
That is, if the models are viewed as having a sufficiently modest
regime of applicability, some of the concerns outlined in this
paper may not apply. There is a second obvious need for caution.
The state of the art in the field is in flux; there is no
generally accepted scheme for how one should implement models of
holographic QCD.  Thus, even if the issues discussed in this paper
do invalidate some aspects  of the various models presently on the
market, it remains quite possible that new variants might be
constructed which will evade them.  With these words of caution,
let us begin our discussion.

The models are holographic in the following sense: it is assumed
that QCD (or some other theory which in some sense approximates
QCD) lives in a four-dimensional world which is the boundary of a
higher-dimensional curved space (typically five dimensional) on
which there exists a dual theory with the property that
correlation functions obtained by varying sources constrained to
the four-dimensional boundary of the dual theory reproduces the
correlation functions of QCD (or the theory which approximates
QCD).  Since all  physical information about a theory is
effectively  contained in its correlation functions one can thus
learn properties of QCD by solving its higher-dimensional dual.
The models are called ``holographic" since information contained
in a lower dimension (in this case the boundary theory) allows one
to see what is going on in a higher dimension in much the same way
that the information in a two-dimensional hologram allows one to
see three-dimensional images.

Now, as stated, this construction is  generically of little use
for studying QCD: if one is replacing one quantum field theory
(QCD) that you cannot solve by a higher-dimensional quantum field
theory which you also cannot solve, you have made no progress.  On
the other hand, suppose that there is a formal limit for which the
higher-dimensional theory becomes weakly coupled and, hence,
classical while the four-dimensional theory remains a strongly
coupled quantum theory.  Then, to the extent one is near this
limit, one can solve the higher-dimensional theory classically to
learn about the quantum physics of the lower-dimensional theory.
All models of the bottom-up holographic QCD are based either
explicitly or implicitly on the assumption that there exists a
five-dimension dual QCD and  for that reason the theory is near
such a limit allowing a classical treatment of the
five-dimensional theory.  I will refer to such a model as
``practical'' since if this condition is false then the
five-dimensional theories are not likely to be tractable in
practice.

This raises obvious questions: Does a practical holographic dual
for QCD exist? If so, what limit of QCD makes the dual theory
classical?  If a practical holographic dual does not exist, to
what extent can practical holographic models mimic key aspects of
QCD?

The hope that such a practical theory exists is based on  the
conjectured dualities between some types of gauge theories and
higher-dimensional gravity theories---most notably the AdS/CFT
correspondence \cite{Maldacena,GubserEtAl,Witten}.  In this
correspondence, a four-dimension conformal field theory, such as
${\cal N} = 4$ $SU(N_c)$ super Yang-Mills theory (in the large
$N_c$ limit), is dual to a type IIB string theory on
$AdS_5\times\mathbf{S}^5$;  $AdS_5$ is 5D Anti-deSitter space, and
$\mathbf{S}^5$ is the 5-sphere. The remarkable thing is then when
the  't Hooft coupling of the conformal field theory is large, the
$AdS_5\times\mathbf{S}^5$ physics is described by weakly-curved
classical supergravity.  Moreover, the construction is explicitly
holographic: the CFT lives on the boundary of $AdS_5$; each
operator $\mathcal{O}$ of the CFT is identified with a bulk field
in $AdS_5\times\mathbf{S}^5$ according to a standard
``dictionary''; the sources for the CFT operators are taken to be
the boundary values of the bulk fields; and the supergravity
partition function $Z_{\mathrm{SG}}$ is identified as the
generating functional of the CFT correlation functions.  If one
denotes the bulk field associated with an operator $\mathcal{O}$
as $\phi^{\mathcal{O}}$, then correlation functions in the CFT are
given by
\begin{equation}
\label{ActionMatching}
Z_{\textrm{SG}}[\phi^{\mathcal{O}}_{0}]=\int_{\phi^{\mathcal{O}} \rightarrow \phi^{\mathcal{O}}_{0}}{D\phi^{\mathcal{O}}\: e^{-S_{SG}[\phi^{\mathcal{O}}] }}
=\left< e^{-\int_{\partial \mathrm{AdS}}{\phi^{\mathcal{O}}_{0} \mathcal{O}}} \right>_{\mathrm{CFT}} .
\end{equation}
The remarkable aspect of this construction is the strong-weak
duality.  The regime where the gauge theory is strongly coupled
corresponds to the weakly coupled 5D theory.  Thus, the CFT
correlation functions in the strongly-coupled regime are obtained
from the  supergravity action $S_{SG}$ evaluated with the {\it
classical} solution for fields $\phi^{\mathcal{O}}$; these
approach  a specified boundary value $\phi^{\mathcal{O}}_0$, and
appropriate functional derivatives with respect to
$\phi^{\mathcal{O}}_0$ are taken.\cite{GubserEtAl,Witten}.

The AdS/CFT correspondence shows that at least some gauge theories
exist which have practical higher-dimensional dual theories in the
sense that they can be treated classically.  Being gauge theories,
they are relatives of QCD.  The hope is that QCD can be treated
similarly.  Note that in constructing the practical dual theory
for the CFT, two limits were critical: the large $N_c$ limit
(which reduced the theory to SUGRA from a string theory) and the
strong coupling limit (which renders the five-dimensional theory
classical.  Presumably some analog of these will be needed for
holographic models of QCD. Thus one expects that a viable
treatment of QCD will require the QCD to be in the large $N_c$
limit and the strongly coupled regimes.  The large $N_c$ limit is
straightforward intellectually---but has some important
implications when attempts to do phenomenology.  The strong
coupling regime is more problematic and will be discussed below.

The bottom-up approach to holographic QCD involves the following
basic steps: guessing a 5D background; taking a field content that
captures some aspects of large $N_c$ QCD for some observables of
interest; a dictionary is used to relate the QCD operators to the
bulk fields (this is generally taken to be the AdS/CFT dictionary
using the naive dimensions of the QCD operators); guessing a form
for the action of the five-dimensional theory; in building this
action one typically uses the 5D masses obtained from the AdS/CFT
dictionary.  Once these steps are taken the problem becomes
computational.  This is called a ``bottom-up'' approach to
distinguish it from a ``top-down'' philosophy in which one
attempts to construct the dual theory from string theory rather
than guess its form.

Some of the guesses in this game are easy to motivate. In AdS/CFT,
the conformal invariance on the 4D side reflects coordinate
rescaling as an isometry of AdS. Now there is a sense in which QCD
is effectively conformal at high energies: due to asymptotic
freedom at high momentum  there is no scale remaining and things
act conformally.  However, at low momenta confinement sets a
natural scale and effective conformal invariance is lost.  Since
confinement is critical to hadronic physics, an unmodified $AdS_5$
background can not capture the essence of QCD; minimally $AdS_5$
space must be modified to reflect confinement.  Typically, the
form guessed for the background in holographic QCD is
asymptotically $AdS_5$ background near the UV-brane; this captured
the effectively conformal nature of QCD at high energies. The deep
bulk region---corresponding to IR physics---is then taken to
deviate from AdS in order to model confinement.  A popular simple
choice is a hard
wall\cite{EKSS,DRP1,DRP2,KatzTensorMesons,HirnRuizSanz,ErlichKribsLow,ShockWu,Forkel},
in which confinement is modeled  by cutting off the AdS space at
some finite radius by hand.  The simplicity of such a model comes
at phenomenological cost: the models do not produce a linear Regge
meson mass spectrum as expected in QCD.   Soft wall models cure
this. They use a dilaton field in the bulk which is tuned to
effectively cut off the AdS space in a smmoth way that is designed to reproduce the
Regge spectrum \cite{KKSS,Forkel,KweeLebed,ColangeloEtAl}. There
are alternative approaches which models include the back-reaction
of the bulk fields on the metric, which can dynamically cut off
the AdS space \cite{CsakiReece,ShockBackreacted}.

There are a number of issues which threaten the viability of such
an approach. Perhaps the most obvious one is the reliance on a
formalism taken over from AdS/CFT which depends on the strong
coupling limit of the CFT.  Now in a CFT, the coupling does not
run and it is thus meaningful to ask whether the theory is
strongly coupled.  In QCD the coupling runs so it is difficult to
even formulate the question of whether the theory is strongly
coupled. To the extent the question can be posed it must be posed
on a process-by-process basis.  This is particularly worrying
since at high momenta where QCD looks effectively conformal it is
weakly coupled.  Thus generically QCD, there is no region in which
QCD is both very strongly coupled and effectively conformal.
Accordingly it is highly questionable whether there is any region
of QCD where it is legitimate to expropriate any aspects of the
AdS/CFT formalism for this system.

There are a large number of subsidiary issues related to this.
From the beginning many of the models fix parameters by matching
to QCD in the UV\cite{EKSS,DRP1,DRP2}.  In part this is done to
limit the number of parameters so the model can have more
predictive power and in part since QCD is tractable in this
regime.  However, this practice makes manifest the problem
discussed above.  It is based on matching to QCD in its
weakly-coupling regime; yet the expectation is that the AdS/CFT
construction which motivates the model is only valid for strong
coupling.

Another conceptual issue with the implementation of the models
concerns the AdS/CFT dictionary\cite{GubserEtAl,Witten} which is
typically used without modification in the bottom-up approach. The
bulk field content is determined by associating a p-form 4D QFT
operator $\mathcal{O}$ with scaling dimension $\Delta$ to a p-form
bulk field with the five-D mass $m_{5}$ uniquely specified by $\Delta$ and $p$.
The not so
subtle issue here is that the operators in QCD run due to their
anomalous dimensions---except for the case of conserved currents.
This raises the obvious question of whether it is sensible to use
the naive dimension of the operator in implementing the
dictionary.  As noted in ref.~\cite{CCW} this also raises
phenomenological theoretical issues in fitting parameters
associated with these operators unless the models can be somehow
augmented to match the scale dependence known in QCD.

A related issue concerns the set of QCD operators---and
corresponding five-dimensional fields---included in the modeling.
QCD has an infinite number of operators with any set of quantum
numbers.  In general these operators mix: the cross correlation
functions are non-zero.  Moreover there is no obvious suppression
scale.  Thus, in principle, an arbitrarily large subset of them
can contribute to any given process\cite{Glozman}.  An {\it ad
hoc} approximation is typically made in order to construct
tractable holographic models: the models typically are restricted
to a minimal set of lowest-dimension operators which probe the
quantum numbers of interest.  Apart from its {\it ad hoc} nature
this procedure raises a conceptual problem.  Suppose that one
assumes that in principle the models treated classically ought to
be rich enough to reproduce {\it any} QCD correlation function,
including the nonvanishing mixed correlators  between distinct
operators of the same quantum numbers. Suppose furthermore that
one associates each of these operators with a distinct bulk field
fixed by the AdS-CFT dictionary and use a classical treatment with
some geometrical background and some action. Then in order to
reproduce the QCD level correlators, it is apparent these distinct
fields must interact classically (either in the bulk or on the
brane). At this level there is no conceptual difficulty.  However,
now suppose we wish to obtain a simpler model with only a minimal
set of fields (say one per quantum number); then one needs to
integrate out the bulk field's extra degrees of freedom, yielding
a generally nonlocal five-dimensional theory for the remaining
degrees of freedom\cite{KKSS}.  One might hope to make some type of field
redefinition to yield an approximately local theory with the
remaining fields. This poses a possible problem: one might expect
that the modification of the action due to integrating out the
extra fields and redefining the surviving ones can alter the 5D
mass term in the action away from that given by the AdS/CFT
dictionary.  Thus it is unclear why the mass terms in the simple
models are given by the dictionary.

There is another set of issues with bottom-up holographic models
of QCD which needs to be addressed, namely, the extent to which
models of this type are capable of reproducing the essential
physics of various aspects of QCD.  To make this question a bit
more crisp we should recall that one of the key ingredients
allowing for a simple classical treatment in the AdS/CFT
formulation was the large $N_c$ limit.  It is hard to believe that
an AdS/QCD connection should be less dependent on large $N_c$.
This does not mean that one cannot try to  include some finite
$N_c$ corrections when doing phenomenology.  However, such
corrections will presumably be included in an {\it ad hoc} way.
The significant point here though is that to the extent  a
holographic model is viable for describing some class of
phenomena, it should {\it at least} be able to describe the
phenomena at large $N_c$.  Since qualitatively much is known about
large $N_c$ phenomenology there are important constraints on model
building.

In fact, much of the modeling does depend critically on large
$N_c$.  For example, simple classical holographic models of mesons
yield mesons with well-defined masses.  To the extent widths are
treated, they are included as perturbative corrections to a
would-be stable state.  In general such a treatment is not
justified but it is at large $N_c$ where meson widths
automatically go to zero as $1/N_c$\cite{tHooft,WittenNc}.
However, this does {\it not} apply to baryons which generically
have a width of order $N_c^0$ if decays are kinematically
allowed\cite{WittenNc}.  Thus holographic models of baryons  such
as those in \cite{baryon} which use field operators taken to be
dual to currents with baryon quantum numbers (in analogy to the
treatment of mesons) and which give rise to excited baryons with
well-defined masses appear to be inconsistent with large $N_c$
QCD.  Moreover one cannot evade this problem by simply working at
$N_c=3$ since baryons are not narrow in the $N_c=3$ world either.
This difficulty need not exclude the possibility of describing
baryons with holographic models\cite{baryon-skyrme} if one
describes baryons arising essentially as Skyrmions\cite{Skyrme},
{\it i.e}, as solitons from the non-linear dynamics of the mesons.
However, it does illustrate how the known large $N_c$ behavior
constrains model building.

The question of whether a practical holographic model is capable
of reproducing the  properties of QCD matter at non-zero
temperature is particularly important\cite{FiniteT}.  On the one
hand, the description of  the phase structure of strongly
interacting matter and the problem of QCD at finite temperature
more generally are at the core of modern nuclear physics and is
tied to experimental studies of ultra-relativistic heavy ion
collisions. However, there is a fundamental issue which arises in
describing QCD at finite temperature using a practical holographic
model.  This is associated with a universal feature of holographic
theories with five-dimensional supergravity duals.  Consider a
theory in the limit  in which such theories may be treated
classically on the 5D side (as is implicitly assumed to be true in
practical holographic models of QCD).  Furthermore assume that the
4D theory is in its deconfined phase.  As noted by Witten such a
phase is  equivalent to there being a black hole in the geometry
of the five-dimensional space\cite{WittenT}.  In such geometries
{\it all}  theories in the class under consideration have a ratio
of shear viscosity to entropy density, $\eta/s$ given by $(4
\pi)^{-1}$ \cite{eta}.  This presents a fundamental difficulty in
trying to describe QCD at high temperatures.  We know that as $T$
gets large, $\eta/s$ should diverge in a logarithmic manner with
$T$  due to asymptotic freedom; practical holographic models based
on SUGRA cannot describe this regime.  One might try to evade this
problem by adding higher-order terms to the gravity theory which
allows $\eta/s$ to differ from $(4 \pi)^{-1}$ \cite{etahd}.
However it is by no means clear that there is a viable way to do
this.  An alternative strategy is to suggest that the models are
only valid for  describing  finite temperature QCD up to a value
not too far above the critical temperature $T_c$\cite{FiniteT}.
For this strategy to be viable, $\eta/s$ at $T_c$ in large $N_c$
QCD needs to be $(4 \pi)^{-1}$ or very close.   {\it A priori}
there is no reason to believe that this is true. The phase
transition is first order at large $N_c$ meaning it is driven by
global energetics rather than local properties of the phase.  Thus
nothing special happens in the deconfined phase as one approaches
$T_c$ from above.  Accordingly it is very hard to conceive of any
reason why one would expect $\eta/s$ to approach $(4 \pi)^{-1}$ as
$T$ approaches $T_c$.

The crux of this problem is the one discussed earlier: QCD has
both strongly coupled and weakly coupled aspects.  The high
temperature phase of QCD has weakly coupled quarks and glue and
this is precisely what causes $\eta/s$ to diverge.  On the other
hand, the classical treatment of the 5D models is only expected to
work for strongly coupled theories.  Thus at high temperature
there is a fundamental tension between the dynamics of QCD and the
structure of  practical holographic models.

There is a final set of challenges to consider.  As noted earlier,
at best the practical holographic models of QCD are justified in
the large $N_c$ limit.  To the extent that some $1/N_c$
corrections are included (by, for example, fitting parameters to
real world data) they are not included in any systematic way. The
large $N_c$ limit and the $1/N_c$ expansion are of use in doing
phenomenology only to the extent that the large $N_c$ world is at
least a rough caricature of the $N_c=3$ world.  However, the
extent to which the large $N_c$ world resembles the physical world
is highly dependent on which observable one is interested in
studying.  It is often the case for hadronic observables that the
large $N_c$ limit {\it does} provide a recognizable, if crude, of
the physical world.  However, even in the domain of hadronic
physics there are cases where the standard large $N_c$ treatment
fails phenomenologically.  The most well-known example of this
concerns the $U(1)$ problem\cite{U1}.  Thus, one should view
results associated with the $\eta '$ meson in holographic models
with particular caution.

More importantly as one goes away from the domain of hadronic
physics to issues in nuclear physics as in ref.~\cite{nuc}, the
large $N_c$ limit and, hence, practical holographic models becomes
more problematic.    The difficulty has to do with energy and
momenta scales in nuclear physics which are characteristically
much smaller than in hadron physics.  Characteristic energies and
momenta in hadronic physics are typically hundreds of MeV to a
GeV.  So long as the energetics associated $1/N_c$ effects are
small on this scale it is presumably legitimate to use a $1/N_c$
expansion and truncate it at relatively low order.  A critical
energy scale to keep in mind is the nucleon-$\Delta$ mass
splitting, a $1/N_c$ effect which is numerically $\sim$290 MeV.
The $1/N_c$ expansion for baryons depends on the excitation of
typical baryons to be large compared to this.   In practice this
is somewhat marginal for many baryons and the $1/N_c$ expansion at
low order for such cases typically should be considered to be
semi-quantitative.  However, in the  domain of nuclear physics the
characteristic energy scales are {\it much} smaller. For examples
the binding energy of the deuteron is approximately 2 MeV or the
binding energy per particle of nuclear matter is approximately 16
MeV.  Using Witten's standard analysis\cite{WittenNc}, it is easy
to see that nuclear binding energies are of order $N_c^1$---just
as baryon masses are. Thus, order $N_c^1$ nuclear quantities such
as these binding energies are smaller by more than an order of
magnitude than the nucleon-$\Delta$ splitting which is formally of
order $1/N_c$. This strongly suggests that treatments based on the
leading order of the $1/N_c$,  such as classical treatments of the
Skyrme model or holographic models of QCD, are likely to fail to
describe nuclei accurately.  Configurations which admix $\Delta$
components can be included with no energetic penalty due to the
N-$\Delta$ mass difference in such treatments. Such admixtures
will be induced to maximize attraction. However, repulsive
contributions due the N-$\Delta$ which are formally higher order
in $1/N_c$ are likely to be much larger than the physical binding
energies given the small size of the latter compared to the
former.

The preceding argument strongly suggests that while a treatment of
low energy nuclear  structure based on the leading order $1/N_c$
expansion might be valid for, say, a world with  $N_c=1001$, it is
almost certainly not valid for the real world of $N_c=3$.  This in
turn implies that whatever else practical holographic models of
QCD may be good for, they are very unlikely to be useful
phenomenological models for nuclear structure.  Of course,
holographic models are not special in this regard; {\it any} model
of the strong interactions which relies on the large $N_c$ limit
is unlikely to be capable of describing nuclear structure.

There is another important domain in which reliance on the large
$N_c$ limit implicit in practical holographic models of QCD is
likely to be problematic phenomenologically: the finite
temperature domain near the QCD phase transition.  It is believed
that the QCD phase transition at large $N_c$ is strongly first
order with a latent heat of order $N_c^2$\cite{WittenNc}; this is
consistent with what has been observed on the lattice\cite{Teper}.
On the other hand, QCD with $N_c=3$ and two light but not massless
quarks, there is no phase transition at all but a crossover; as
the quark masses go to zero a second-order phase transition
occurs.  Thus the transition region in the physical world and the
extrapolation of the physical world to the chiral limit looks
qualitatively quite different than the large $N_c$ limit.  This is
a problem because any attempt to describe this physics in a
practical holographic model will necessarily depend on
uncontrolled higher-order corrections in a $1/N_c$ expansion.  An
optimistic view is that the large $N_c$ limit could still
accurately describe the system except in a narrow region around
the phase transition.  However, it is worth noting from the
discussion of $\eta /s$ that the models are expected to fail well
above $T_c$ and from the discussion above it seems apparent that
they also are expected to fail near $T_c$.  This raises an obvious
question: Is there {\it any} region in the deconfined phase where
the models do work?

In summary, practical ``bottom-up'' holographic models  of QCD are
quite interesting.  However, they face a number of challenges both
at the theoretical and phenomenological levels.  Before accepting
models of this class as useful descriptions of strong interacting
physics it is important that these be addressed.  It also seems
clear that at best the domain of validity of these models is
likely to be rather limited.

\begin{acknowledgements}
The support of the United States Department of Energy is gratefully acknowledged.  Discussions with A. Cherman were extremely helpful in formulating the arguments in this paper.
\end{acknowledgements}

\end{document}